\documentclass[letterpaper,english,reprint, prl]{revtex4-2}
\usepackage{lmodern}

\usepackage[T1]{fontenc}
\usepackage[latin9]{inputenc}
\usepackage{color}
\usepackage{babel}
\usepackage{verbatim}
\usepackage{amsmath}
\usepackage{amssymb}
\usepackage{graphicx}
\usepackage[unicode=true,pdfusetitle,
 bookmarks=true,bookmarksnumbered=false,bookmarksopen=false,
 breaklinks=false,pdfborder={0 0 1},backref=false,colorlinks=true]
 {hyperref}
\hypersetup{
 linkcolor=blue, urlcolor=blue, citecolor=blue}

\makeatletter

\pdfpageheight\paperheight
\pdfpagewidth\paperwidth

\usepackage{hyperref}
\usepackage[all]{hypcap}

\makeatother

\begin{document}
\title{Aging by near-extinctions in many-variable interacting populations}
\author{Thibaut Arnoulx de Pirey and Guy Bunin}
\affiliation{Department of Physics, Technion-Israel Institute of Technology, Haifa
32000, Israel}
\begin{abstract}
Models of many-species ecosystems, such as the Lotka-Volterra and
replicator equations, suggest that these systems generically exhibit
near-extinction processes, where population sizes go very close to
zero for some time before rebounding, accompanied by a slowdown of
the dynamics (aging). Here, we investigate the connection between
near-extinction and aging by introducing an exactly solvable many-variable
model, where the time derivative of each population size vanishes
both at zero and some finite maximal size. We show that aging emerges
generically when random interactions are taken between populations.
Population sizes remain exponentially close (in time) to the absorbing
values for extended periods of time, with rapid transitions between
these two values. The mechanism for aging is different from the one
at play in usual glassy systems: at long times, the system evolves
in the vicinity of unstable fixed points rather than marginal ones.
\end{abstract}
\maketitle
Interactions between species in ecosystems may lead to large fluctuations
in their population sizes. Theoretical models play a central role
in understanding these fluctuations in nature and experiments, both
for several species \citep{beninca2015species,fussmann2000crossing,gause_volterra,venturelli2018deciphering}
and for many species \citep{hu2022emergent}. The dynamics of populations
that interact and reproduce are often modeled by coupled ordinary
differential equations for population sizes $\left\{ x_{i}\right\} $.
They are non-negative variables, $x_{i}\ge0$, and must remain so
throughout the dynamics. The boundary values $x_{i}\left(t\right)=0$
represent extinct populations: if a population is extinct at some
time $t$, it must remain so at all later times. Namely, $x_{i}=0$
is an absorbing value for $x_{i}$. These requirements are satisfied
by a broad class of differential equations of the form \citep{hofbauer_evolutionary_1998}

\begin{equation}
\dot{x}_{i}=x_{i}g_{i}(\vec{x})\,.\label{eq:ecol_diff}
\end{equation}
Examples in this class include the Lotka-Volterra equations for which
$g_{i}(\vec{x})=B_{i}-\sum_{j}A_{ij}x_{j}$, with the matrix $\mathbf{A}$
encoding the interactions between populations; resource-competition
models \citep{arthur_species_1969}; and the replicator equations
employed in evolution and game theory \citep{hofbauer_evolutionary_1998}.

\begin{figure}
\begin{centering}
\includegraphics[width=1\columnwidth]{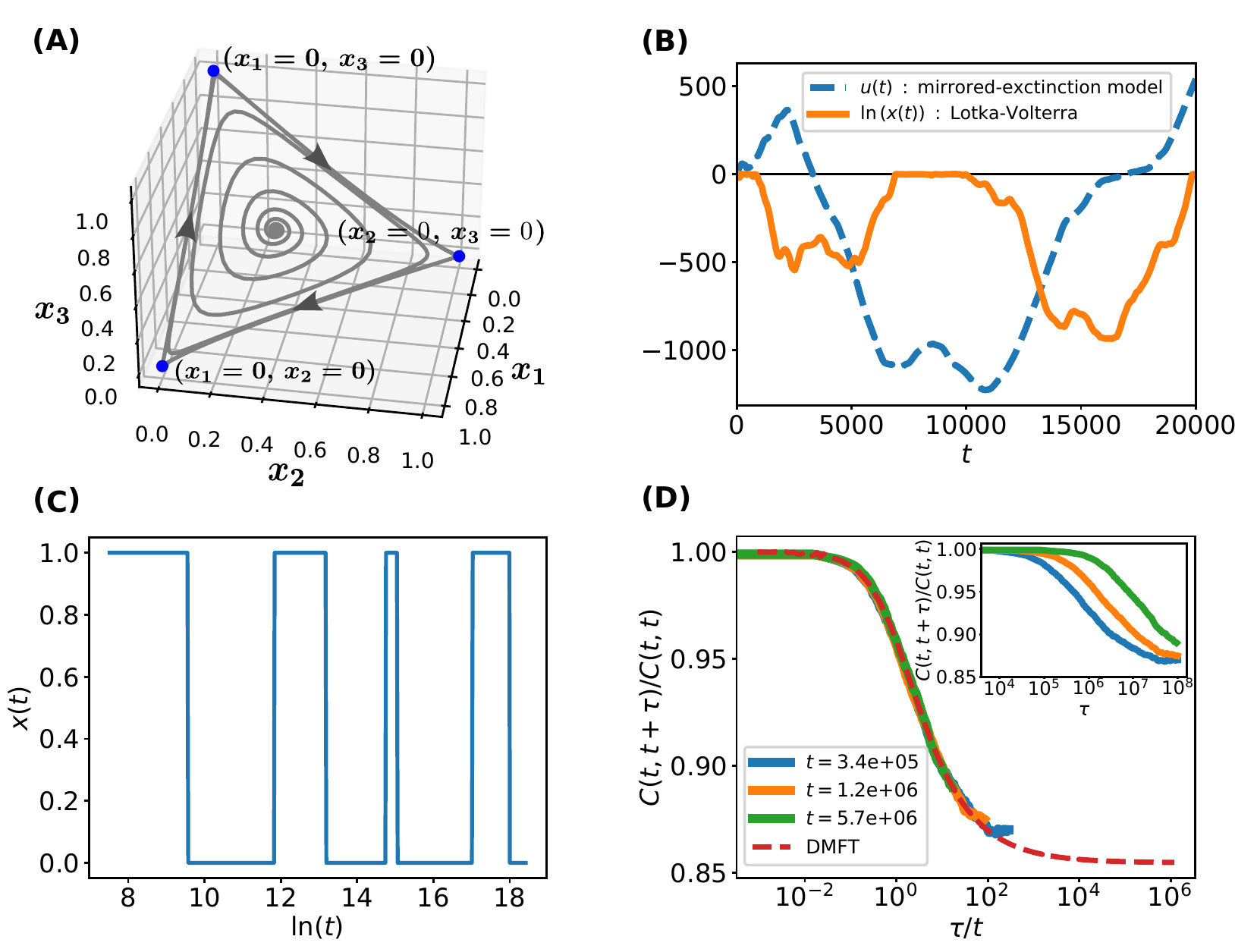}
\par\end{centering}
\caption{\label{fig:invadability and cycles non-jumps}\textbf{Aging by passing
near unstable fixed points.} \textbf{(A)} Heteroclinic cycle in the
three-variable May-Leonard model. The dynamics slow down as the system
goes ever closer to fixed points (dots), despite them being unstable.
\textbf{(B)} Dynamics of example variables (out of $S=2\cdot10^{4}$),
in the Lotka-Volterra system (solid line, plot of $\ln x_{i}$) and
the mirrored-extinction model (\ref{eq:x_model-1}), (dashed, plot
of $\ln[x_{i}/(1-x_{i})]$). This illustrates the longer and deeper
excursions near the absorbing values, $x_{i}=0$ for Lotka-Volterra
and $x_{i}\in\left\{ 0,1\right\} $ for the mirrored-extinction case.\textbf{
(C)} In log-time, the dynamics of any variable in (\ref{eq:x_model-1})
eventually follow a biased time-translation invariant two-state process.\textbf{
(D)} Mean autocorrelation function $C(t'+\tau,t')$ of $x_{i}(t)$
as measured in a numerical simulation of (\ref{eq:x_model-1}) with
$2\cdot10^{4}$ degrees of freedom as a function of $\tau/t'$, showing
a collapse for different waiting times $t'$, and agreement with the
analytical master curve (dashed line). Inset: same curves, as a function
of $\tau$. Parameters for Lotka-Volterra simulations in (A,B) are
given in App.~\ref{sec:Lotka-Volterra-simulations}.}
\end{figure}

It is well-known that, depending on the shape of the functions $g_{i}$,
few variable systems of the form (\ref{eq:ecol_diff}) can exhibit
different long-time behaviors such as stationarity, periodicity or
chaos \citep{hofbauer_evolutionary_1998}. Remarkably, the existence
of absorbing hyperplanes has also been shown to lead, in some cases,
to robust heteroclinic cycles \citep{krupa_robust_nodate,hofbauer_heteroclinic_1994}.
A classical example is the three-species Lotka-Volterra system with
rock-paper-scissors type interactions \citep{may_nonlinear_1975},
where each species hinders the growth of the next. There, trajectories
are attracted to a cycle connecting three unstable fixed points, each
with a single surviving population, see Fig. \ref{fig:invadability and cycles non-jumps}(a).
As time increases, they pass ever closer to these fixed points, resulting
in slowdown of the dynamics, with exponentially increasing sojourn
times in their vicinity and rapid transitions between them \citep{,gaunersdorfer_time_1992-1,may_nonlinear_1975}.

In models characterized by a large number $S$ of variables, recent
works find that an analogous slowdown emerges generically for random
interaction coefficients. It is known that aging (a situation in which
the system does not asymptotically settle to a fixed point but keeps
exploring the phase space with a velocity that nevertheless decays
with the elapsed time) can occur in many-variable Lotka-Volterra systems
with random asymmetric interactions \citep{roy_numerical_2019} and
replicator equations with nearly antisymmetric random interactions
\citep{pearce_stabilization_2020}. Here, some populations experience
ever longer periods near extinction ($x_{i}\simeq0$ and $x_{i}$
closer to zero in successive near-extinction periods) before eventually
returning to $x_{i}=O(1)$, see Fig. \ref{fig:invadability and cycles non-jumps}(b).
Such dips and `blooms' are documented in experiments and field data
(e.g, \citep{martin-platero_high_2018,ignacio-espinoza_long-term_2020}),
and are ecologically significant as they may lead to extinctions in
actual finite populations. The properties of these dynamics have remained
elusive, however. The analogy with low-dimensional examples such as
in Fig. \ref{fig:invadability and cycles non-jumps}(a) is limited.
For one, in the many-variable case, the system does not approach a
limit-cycle (at least if the limit $S\to\infty$ is taken before $t\to\infty$).
Secondly, large dynamical systems of the form (\ref{eq:ecol_diff})
may possess many fixed points with different properties (e.g., the
fraction of variables for which $x_{i}=0$ or their instability index),
and linking the characteristics of fixed points to the dynamics remains
an open problem.

In this work, we propose a high-dimensional model of the form (\ref{eq:ecol_diff})
that provides insights into the connection between aging and absorbing
values, by bypassing some of the difficulties inherent to the many-variable
Lotka-Volterra and replicator equations. Fixed points of (\ref{eq:ecol_diff})
satisfy either $x_{i}=0$ or $g_{i}(\vec{x})=0$ for every $i$. Since
the unique aging behavior of these systems is tied to the existence
of absorbing values ($x_{i}=0$), we introduce a model with \emph{two}
absorbing values for each variable, which we refer to as the mirrored-extinction
model. Specifically, we consider the evolution of $S$ degrees of
freedom $\{x_{i}\}_{i=1,\ldots,S}$, with $0\le x_{i}\le1$ for all
$i$,

\begin{equation}
\dot{x}_{i}(t)=x_{i}(t)\left[1-x_{i}(t)\right]\sum_{j=1}^{S}\alpha_{ij}x_{j}(t)\,,\label{eq:x_model-1}
\end{equation}
where $\mathbf{\boldsymbol{\alpha}}$ is a zero-mean Gaussian random
matrix with independent and identically distributed entries (referred
to as asymmetric interactions). We take $\mathbb{E}\left[\alpha_{ij}^{2}\right]=1/S$
which sets the units of time. From an ecological perspective, the
interactions in (\ref{eq:x_model-1}) affect the growth-rates of populations
but not their maximal size, which might be limited by other factors,
see for example \citep{ratzke_modifying_2018,ratzke_strength_2020}.
Equation (\ref{eq:x_model-1}) can be extended by adding a species-dependent
growth rate $g_{i}$ to the sum, $\sum_{j}\alpha_{ij}x_{j}\rightarrow g_{i}+\sum_{j}\alpha_{ij}x_{j}$
(so that when a species is alone it undergoes simple logistic growth
with a growth rate $g_{i}$, similarly to the Lotka-Volterra equations),
and is solvable just as described below and with the same qualitative
outcomes, see App.~\ref{sec:Growth-Rate}.

The resulting dynamical system has many fixed points where all degrees
of freedom are at their absorbing values, either $x_{i}=0$ or $x_{i}=1$,
allowing us to focus on the effects of these absorbing boundaries.
It displays aging, similarly to the Lotka-Volterra case, but with
the $x_{i}$ spending ever longer times close to either $x_{i}=0$
or $1$ with rapid transitions between these two values, see Fig.
\ref{fig:invadability and cycles non-jumps}(b). Importantly, the
model in (\ref{eq:x_model-1}) is exactly solvable in high dimension,
allowing us to obtain detailed information on the link between near-extinction
processes and aging, beyond other models that also exhibit both phenomena
\citep{roy_numerical_2019,pearce_stabilization_2020}.

The mechanism for aging found here is drastically different from that
at play in aging of usual spin-glasses following a quench, where the
system's energy is reduced until it reaches an energy surface dominated
by marginally-stable fixed points and spends its time there \citep{cugliandolo_analytical_1993,kurchan_phase_1996,manacorda_gradient_2022}.
This includes Lotka-Volterra dynamics with \emph{symmetric} interaction
matrices $\alpha_{ij}$ \citep{altieri_properties_2021,biroli_marginally_2018},
where $\vec{g}(\vec{x})$ is the gradient of a potential, thus permitting
a mapping to a spin-glass phase. This form of aging is known to disappear
when asymmetry is introduced to the interaction coefficients \citep{cugliandolo_glassy_1997,crisanti_dynamics_1987}.

In contrast, here we show that aging happens in (\ref{eq:x_model-1}),
as variables are driven close to their absorbing values: the probability
$P(x_{i})$ at long times concentrates about $\left\{ 0,1\right\} $,
as shown below in (\ref{eq:steadystate}). Near fixed points the dynamics
slow down, as manifested in the autocorrelation $C(t'+\tau,t')$ of
$x_{i}(t)$, which as $t'$ grows, relaxes more slowly with $\tau$,
as shown below in (\ref{eq:defchat-1}). Similarly to the three-variable
example of Fig. \ref{fig:invadability and cycles non-jumps}(a), typical
systems go very close to fixed points which are therefore long-lived,
see Fig. \ref{fig:invadability and cycles non-jumps}(b,c). This happens
despite these fixed points being \emph{unstable}, which we show by
calculating the spectrum, Eq. (\ref{eq:spectrum_analytical}), of
the linearized dynamics around the fixed points approached at long
times. This provides a mechanism for aging in the absence of an underlying
energy function. We find that, in the long-time limit, the system
moves between infinitely many unstable fixed points that all have
the same finite fraction of unstable directions and the same stability
spectrum. They are neither the most stable nor the most abundant fixed
points.

\begin{comment}
To conclude, we propose an exactly-solvable many-variable model for
the dynamics of interacting populations with absorbing boundary values.
Its dynamics slow down with a correlation time that grows as the age
of the system, see (\ref{eq:defchat-1}). The system evolves in the
vicinity of fixed points: In the long-time limit, all variables are
found exponentially close in time to absorbing values, see (\ref{eq:expclose}).
The time it takes for a variable to leave the vicinity of one absorbing
value to visit the vicinity of the other is therefore proportional
to the age of the system. This explains the scaling of the aging,
(\ref{eq:defchat-1}). All these fixed points are unstable, as shown
in (\ref{eq:spectrum_analytical}), in contrast with marginal fixed
points reached in usual glassy dynamics \citep{cugliandolo_analytical_1993,kurchan_phase_1996,manacorda_gradient_2022}.
In the future, it would be interesting to understand how this scenario
is adapted to other many-variable interacting population dynamics,
such as the Lotka-Volterra model, where fixed points have degrees
of freedom that are not at absorbing values. Fingerprints of these
phenomena might be observed, as an increase in correlation time combined
with population blooms, in experiments that follow interacting species
starting from similar population sizes.
\end{comment}

\emph{Dynamical mean field theory\textemdash }To analyze the many-variable
dynamics (\ref{eq:x_model-1}), we use dynamical mean field theory
(DMFT) \citep{mezard_spin_1987,sompolinsky_relaxational_1982}. In
the limit $S\to\infty$ and for $x_{i}$ sampled independently at
the initial time, the dynamics of a single degree of freedom $x\left(t\right)$
are exactly described by a stochastic differential equation

\begin{equation}
\dot{x}(t)=x(t)\left[1-x(t)\right]\xi(t)\,,\label{eq:DMFTx}
\end{equation}
with $\xi(t)$ a zero mean Gaussian process. This stems from the fact
that the term $\xi_{i}(t)\equiv\sum_{j}\alpha_{ij}x_{j}(t)$ appearing
in (\ref{eq:x_model-1}) is the sum of many weakly correlated contributions.
As is usual in DMFT, this expression for $\xi_{i}\left(t\right)$
yields a self-consistent closure relation that reads $C(t,t')\equiv\langle\xi(t)\xi(t')\rangle=\langle x(t)x(t')\rangle$.
Here the angular brackets $\langle.\rangle$ denote an average over
the initial conditions $x(0)$ and realizations of the noise $\xi(t).$
The derivation of (\ref{eq:DMFTx}) follows a standard procedure \citep{mezard_spin_1987,sompolinsky_relaxational_1982,liu_dynamics_2021,roy_numerical_2019,agoritsas_out--equilibrium_2018}
and is detailed in App.~\ref{sec:Dynamical-mean-field}. To analyze the dynamics, it is therefore
very helpful to solve for the autocorrelation function $C(t,t')$.

To proceed, we introduce the transformation $u(t)=\ln\left[x(t)/\left(1-x(t)\right)\right]$
that sends the boundaries of the domain $[0,1]$ to $\left(-\infty,+\infty\right)$
and for which (\ref{eq:DMFTx}) becomes

\begin{equation}
\dot{u}(t)=\xi(t)\,,\label{eq:u_model}
\end{equation}
with the closure relation
\begin{equation}
\langle\xi(t)\xi(t')\rangle=\left\langle f\left(u\left(t\right)\right)f\left(u\left(t'\right)\right)\right\rangle \,,\label{eq:correl_u}
\end{equation}
where
\[
f\left(y\right)\equiv\frac{\mathrm{e}^{y}}{1+\mathrm{e}^{y}}\ .
\]

\emph{Aging and the auto-correlation function\textemdash }We start
by showing that the mean-square displacement of $u(t)$ is ballistic.
Denote the auto-correlation $G(t,t')\equiv\left\langle u(t)u(t')\right\rangle $,
which by (\ref{eq:u_model}) is related to $C(t,t')$ by $C(t,t')=\partial_{t}\partial_{t'}G(t,t')$.
We take initial conditions such that $u(0)=0$, or equivalently $x(0)=1/2$;
the long-time behavior of the correlation function is insensitive
to this choice. The closure equation in (\ref{eq:correl_u}) can then
be written as 
\begin{align}
\partial_{t}\partial_{t'}G(t,t') & =\left\langle f\left(u\left(t\right)\right)f\left(u\left(t'\right)\right)\right\rangle \ .\label{eq:self_cons-1}
\end{align}
$u\left(t\right),u\left(t'\right)$ are jointly Gaussian with zero
means, from which it follows that $1/16\le\left\langle f\left(u\left(t\right)\right)f\left(u\left(t'\right)\right)\right\rangle \le1$,
see App.~\ref{sec:Bounds-on-the}. Therefore $tt'/16<G(t,t')<tt'$, so $\left\langle u(t)^{2}\right\rangle =G(t,t)\sim t^{2}$,
corresponding to \emph{ballistic} growth of $u(t)$. We show below
that $u(t)$ nonetheless repeatedly crosses the origin at arbitrarily
long times.

The long-time expression for $G(t,t')$ can be worked out from (\ref{eq:self_cons-1}).
Here we present a different but equivalent derivation, which makes
explicit the aging properties of the model. Motivated by the ballistic
growth of $u\left(t\right)$, we introduce $z(t)\equiv u(t)/t$, and
we rescale time though $s\equiv\ln(t).$ The resulting dynamics read

\begin{equation}
z'(s)=-z(s)+\hat{\xi}(s)\,,\label{eq:mapp_sompo-1}
\end{equation}
together with the closure relation (from (\ref{eq:correl_u}))
\[
\left\langle \hat{\xi}(s)\hat{\xi}(s')\right\rangle =\left\langle f\left(\text{e}^{s}z(s)\right)f\left(\text{e}^{s'}z(s')\right)\right\rangle \,.
\]
Because $z(s)$ is a Gaussian process with finite $O(1)$ variance
as $s\to\infty$, in the long-time limit this equation reads

\begin{equation}
\left\langle \hat{\xi}(s)\hat{\xi}(s')\right\rangle =\left\langle \Theta(z(s))\Theta(z(s'))\right\rangle \,.\label{eq:lim_closure-1}
\end{equation}
Equations (\ref{eq:mapp_sompo-1},\ref{eq:lim_closure-1}) map the
original many-body dynamics of (\ref{eq:x_model-1}), in the long-time
limit, to chaotic dynamics of random neural networks of the form discussed
in \citep{sompolinsky_chaos_1988}. As in \citep{sompolinsky_chaos_1988},
at large $s$, we expect the process in (\ref{eq:mapp_sompo-1}) to
reach a time-translation invariant chaotic state characterized by

\[
\left\langle \hat{\xi}(s)\hat{\xi}(s')\right\rangle \equiv\hat{C}(s-s')\,.
\]
In the original time scale $t=\text{e}^{s}$, this corresponds to
 autocorrelation of the form,

\begin{equation}
\lim_{t'\to\infty}C(t'+\tau,t')=\hat{C}\left(\ln\left(1+\beta\right)\right)\,,\label{eq:defchat-1}
\end{equation}
at fixed $\beta\equiv\tau/t'$. $C(t'+\tau,t')$ thus relaxes more
slowly with $\tau$ as $t'$ grows, a hallmark of \emph{aging}, here
with correlation time growing linearly with the elapsed time. Accordingly,
from (\ref{eq:mapp_sompo-1}), the $z(s)$ autocorrelation function
also admits a time-translation invariant form at large times

\[
\left\langle z(s)z(s')\right\rangle \equiv\hat{\Delta}(s-s')\,,
\]
which is related to $G(t,t')$ through $\lim_{t'\to\infty}G(t'+\tau,t')/t'(t'+\tau)=\hat{\Delta}\left(\ln\left(1+\beta\right)\right).$
We now sketch the derivation of $\hat{\Delta}$. Following \citep{sompolinsky_chaos_1988},
$\hat{\Delta}$ and $\hat{C}$ are related by $\hat{C}(s)=-\hat{\Delta}''(s)+\hat{\Delta}(s)$
which, together with (\ref{eq:lim_closure-1}), implies that $\hat{\Delta}(s)$
satisfies an equation for the motion of a classical particle in a
potential $V$

\begin{equation}
\hat{\Delta}''(s)=-V'(\hat{\Delta},\Delta_{0})\,,\label{eq:delta_eq-1}
\end{equation}
where the potential depends parametrically on the initial condition
$\Delta_{0}\equiv\hat{\Delta}(0)$ and reads,

\[
V\equiv-\frac{\hat{\Delta}^{2}}{2}+\frac{\hat{\Delta}}{4}+\frac{\hat{\Delta}}{2\pi}\!\left(\sqrt{\frac{\Delta_{0}^{2}}{\hat{\Delta}^{2}}-1}+{\rm arccot}\sqrt{\frac{\Delta_{0}^{2}}{\hat{\Delta}^{2}}-1}\right)\,.
\]
The condition $\hat{\Delta}(s)=\hat{\Delta}(-s)$ implies $\hat{\Delta}'(0)=0$
so that the $\hat{\Delta}(s)$ trajectory has zero initial kinetic
energy. The only physically relevant trajectory is therefore the one
converging to the unstable fixed point $\Delta^{*}$ with same potential
energy as the initial condition and related to $\Delta_{0}$ by $V'(\Delta^{*},\Delta_{0})=0$
and $V(\Delta^{*},\Delta_{0})=V(\Delta_{0},\Delta_{0})$. This gives
$\Delta_{0}\simeq0.476$ and $\Delta^{*}\simeq0.427$.

The correlation $C(t'+\tau,t')=\frac{1}{S}\sum_{i}x_{i}(t'+\tau)x_{i}(t')$,
obtained by running the dynamics (\ref{eq:x_model-1}), is thus expected
by (\ref{eq:defchat-1}) to collapse when plotted against $\tau/t'$,
as indeed seen in Fig. \ref{fig:invadability and cycles non-jumps}(d),
and it matches the correlation function $\hat{C}(s)$ obtained by
numerically solving (\ref{eq:delta_eq-1}) with the appropriate initial
conditions. Note that $\Delta_{0}$ is linked to the long-time growth
of $\ensuremath{\left\langle u(t)^{2}\right\rangle }$, as $G(t,t)/t^{2}\underset{t\to\infty}{\to}\Delta_{0}$.
Additionally, the auto-correlation satisfies
\begin{equation}
\lim_{t'\to\infty}C(t',t')=\frac{1}{2}>\lim_{\tau\to\infty}\lim_{t'\to\infty}C(t'+\tau,t')=\Delta^{*}\,,\label{eq:correl_limit}
\end{equation}
so that the system continues to evolve, as the correlation with the
state at any time is later partially lost. Equation (\ref{eq:delta_eq-1})
implies a power law relaxation of $C(t,t')$ in the aging regime to
its plateau value $\Delta^{*}$, 
\[
\lim_{t'\to\infty}C\left(t'(1+\beta),t'\right)-\Delta^{*}\underset{\beta\to\infty}{\sim}\beta^{-k}\,,
\]
with $k=\sqrt{\left|V''(\Delta^{*},\Delta_{0})\right|}\simeq0.492$.

\emph{Single variable dynamics\textemdash }The dynamics (\ref{eq:x_model-1})
pass very close to fixed points at long times. To see this, we calculate
the probability distribution of $x$ at time $t$, $P_{t}\left(x\right)$,
taken over many variables in (\ref{eq:x_model-1}), or equivalently
over different realizations of (\ref{eq:DMFTx}). Using the fact that
$u\left(t\right)$ is Gaussian and that $x(t)=f\left(u\left(t\right)\right)$,
it reads

\begin{equation}
P_{t}(x)=\frac{\left[x(1-x)\right]{}^{-1}}{\sqrt{2\pi G(t,t)}}\exp\left[-\frac{1}{2G(t,t)}\left(\ln\frac{x}{1-x}\right)^{2}\right]\,.\label{eq:pxt}
\end{equation}
In particular this implies,
\begin{equation}
\lim_{t\to\infty}P_{t}(x)=\frac{1}{2}\left[\delta(x)+\delta(x-1)\right]\,.\label{eq:steadystate}
\end{equation}
This shows that the system (\ref{eq:x_model-1}) \emph{asymptotically
approaches fixed points} of the dynamics, where all $x_{i}\in\left\{ 0,1\right\} $.Furthermore,
at large but finite times, the probability to find $x(t)$ away from
the boundaries of $[0,1]$ decays as $1/t$, with (\ref{eq:pxt})
giving

\begin{equation}
\text{Prob}[x(t)\in[\epsilon,1-\epsilon]]\underset{t\to\infty}{\sim}\frac{1}{t}\sqrt{\frac{2}{\pi\Delta_{0}}}\ln\left(\frac{1-\epsilon}{\epsilon}\right)\,.\label{eq:expclose}
\end{equation}
for any fixed $\epsilon\in[0,1/2]$. The probability is thus concentrated
exponentially close in time to 0 and 1. Yet the system continues to
evolve, see (\ref{eq:correl_limit}), so that none of these fixed
points are stable: At long times the system transitions between unstable
fixed points, spending ever longer times in their vicinity with fast
transitions between them.

In the long-time limit, since $x(s)=\Theta(z(s))$ for $s\to\infty$,
$x(s)$ asymptotically approaches a time-translation invariant two-state
process. This is illustrated in Fig. \ref{fig:invadability and cycles non-jumps}(c).
Note that as $\Delta^{*}>0$ in (\ref{eq:correl_limit}), equation
(\ref{eq:steadystate}) is not the ergodic measure (in log-time) of
a single variable $x_{i}\left(t\right)$. In App.~\ref{sec:Ergodic-measure-for}, we show
that for a given degree of freedom the log-time ergodic measure is
given by

\begin{equation}
P_{\bar{\xi}}(x)=(1-p)\,\delta(x)+p\,\delta(x-1)\,,\label{eq:ergodic_measure}
\end{equation}
with $p=\left[1+\text{Erf}\left(\bar{\xi}/\sqrt{2\chi}\right)\right]/2$,
where $\bar{\xi}$ is a zero mean Gaussian random variable with variance
$\left\langle \bar{\xi}^{2}\right\rangle =\Delta^{*}$ and $\chi=\int_{0}^{\infty}ds\text{e}^{-s}\left[\hat{C}(s)-\Delta^{*}\right]$.
So, in a given realization of (\ref{eq:x_model-1}), each variable
has an ``identity'' expressed in the fraction of time (in log-time)
it spends near 0 and 1.

\emph{Stability of visited fixed points\textemdash }We found above
that at long times the system approaches fixed points, but eventually
leaves their vicinity, signaling that they are unstable. We now calculate
their entire stability spectrum. The linearized dynamics close to
a fixed point $\boldsymbol{x^{*}}$ are $\dot{\delta x_{i}}=J_{ij}\,\delta x_{i}$
with a diagonal matrix $J_{ij}=\delta_{ij}\lambda_{i}^{*}$. The growth
rates $\lambda_{i}^{*}$, positive when growing in the direction away
from the boundaries, are given by

\begin{equation}
\lambda_{i}^{*}=(1-2x_{i}^{*})\left(\sum_{j}\alpha_{ij}x_{j}^{*}\right)\,.\label{eq:growth_rates}
\end{equation}
The stability spectrum of the visited fixed points is therefore equal,
at long-times, to the empirical distribution in the many-variable
dynamics (\ref{eq:x_model-1}) of $\lambda_{i}(t)\equiv\left[1-2x_{i}(t)\right]\xi_{i}(t)$
for $i=1\dots S$. In the $S\to\infty$ limit, the stability spectrum
is thus equal to the distribution of $\lambda(t)=\left[1-2x(t)\right]\xi(t)$
in the DMFT framework. It can also be shown that the $\lambda_{i}(t)$
are independent and identically-distributed random variables, see App.~\ref{sec:Stability-spectrum}, therefore the spectrum is self-averaging.

\begin{figure}
\begin{centering}
\includegraphics[width=0.8\columnwidth]{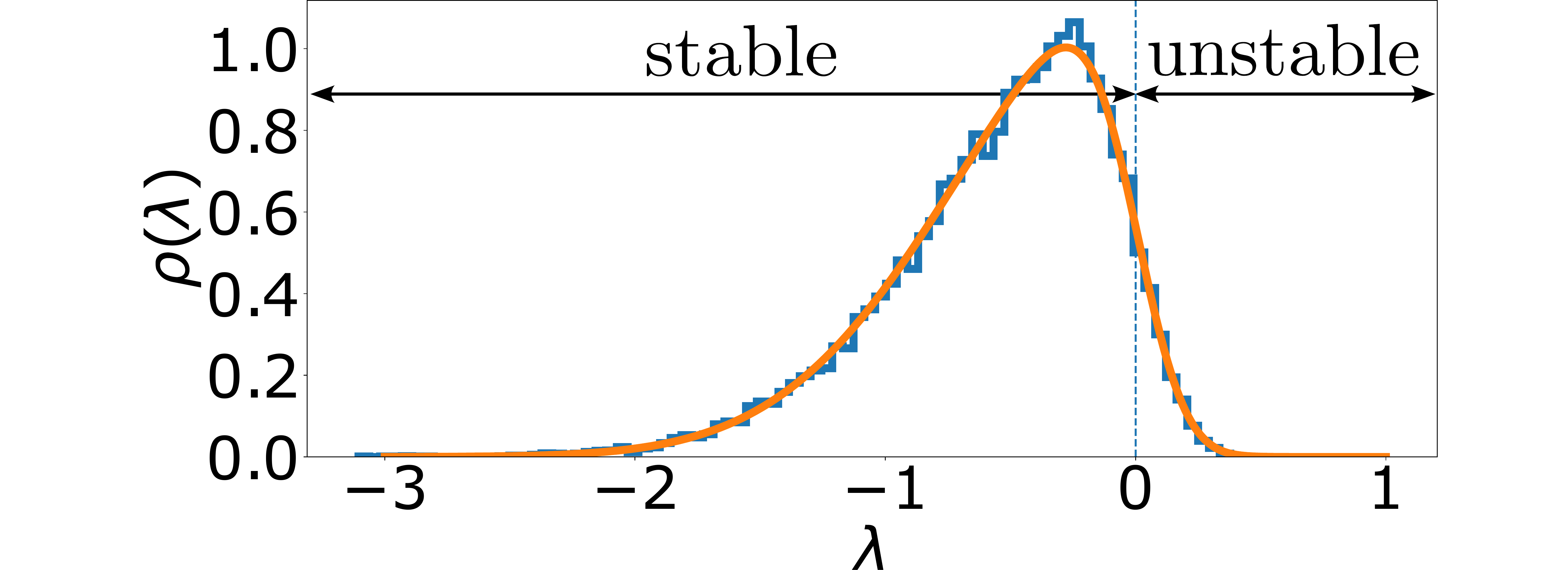}
\par\end{centering}
\caption{\label{fig:spectrum}\textbf{Stability spectrum of the fixed points
visited at long times.} The long-time dynamics evolve in the vicinity
of unstable fixed points which all have the same stability spectrum.
A finite fraction of the eigenvalues are positive, corresponding to
unstable directions around these fixed points. The analytical prediction
for the spectrum (\ref{eq:spectrum_analytical}), is in excellent
agreement with a simulation (blue) with $S=2\cdot10^{4}$ variables
at $t=10^{8}$.}
\end{figure}

The joint distribution of $\xi(t)$ and $u(t)$ is Gaussian, with
correlations $\left\langle u(t)^{2}\right\rangle =G\left(t,t\right)$,
$\left\langle \xi(t)^{2}\right\rangle =C\left(t,t\right)$ and cross-correlation
$\left\langle u\left(t\right)\xi(t)\right\rangle $. Changing variables
from $\left(u,\xi\right)$ to $\left(u,\lambda\right)$ and integrating
over $u$, we obtain an expression for the distribution of $\lambda\left(t\right)$,
reproduced in App.~\ref{sec:Stability-spectrum}. Taking its long-time limit, we find that
the dynamics (\ref{eq:x_model-1}) transition between fixed points
which all have the same stability spectrum

\begin{equation}
\rho(\lambda)=\frac{1}{\sqrt{\pi}}\text{e}^{-\lambda^{2}}\,{\rm Erfc}\left(\frac{\lambda}{\kappa_{\infty}}\right)\,,\label{eq:spectrum_analytical}
\end{equation}
with $\kappa_{\infty}=\sqrt{1/\left(2\Delta_{0}\right)-1}\simeq0.224$,
see Fig. \ref{fig:spectrum}. This distribution has a \emph{finite
fraction of unstable directions}, given by

\[
\int_{0}^{\infty}\rho(\lambda)\,d\lambda=\frac{{\rm arctan(\kappa_{\infty})}}{\pi}\simeq0.141\,.
\]
Thus, the system approaches unstable fixed points. This can be compared
with the statistics of the full distribution of fixed points of (\ref{eq:x_model-1})
with all $x_{i}\in\left\{ 0,1\right\} $. There are $2^{S}$ of them
and the average number of those with $\alpha S$ unstable directions
is given by the binomial law, $\left\langle \mathcal{N}_{\alpha}\right\rangle \sim\exp\left[Sg(\alpha)\right]$,
with $g(\alpha)=-\alpha\ln\alpha-(1-\alpha)\ln(1-\alpha)$. Therefore,
in typical fixed points half of the directions are unstable, $\alpha=1/2$.
The dynamics therefore selects in the long-time limit fixed points
that are exponentially rare (compared to the typical ones) but that
are not the most stable ones existing in the phase space, which are
marginal ($\alpha=0$).

To conclude, we propose an exactly-solvable many-variable model for
the dynamics of interacting populations with absorbing boundary values.
Its dynamics slow down with a correlation time that grows as the age
of the system, see (\ref{eq:defchat-1}). The system evolves in the
vicinity of fixed points: In the long-time limit, all variables are
found exponentially close in time to absorbing values, see (\ref{eq:expclose}).
The time it takes for a variable to leave the vicinity of one absorbing
value to visit the vicinity of the other is therefore proportional
to the age of the system. This explains the scaling of the aging,
(\ref{eq:defchat-1}). All these fixed points are unstable, as shown
in (\ref{eq:spectrum_analytical}), in contrast with marginal fixed
points reached in usual glassy dynamics \citep{cugliandolo_analytical_1993,kurchan_phase_1996,manacorda_gradient_2022}.
In the future, it would be interesting to understand how this scenario
is adapted to other many-variable interacting population dynamics,
such as the Lotka-Volterra model, where fixed points have degrees
of freedom that are not at absorbing values. Fingerprints of these
phenomena might be observed, as an increase in correlation time combined
with population blooms, in experiments that follow interacting species
starting from similar population sizes.

\emph{Acknowledgments}\textemdash G. B. was supported by the Israel
Science Foundation (ISF) Grant No. 773/18.

\bibliographystyle{unsrt}
\bibliography{My_Library}
\clearpage{}
\begin{center}
\onecolumngrid
\appendix
\setcounter{page}{1}\textbf{Supplemental material for \textquotedblleft Aging by near-extinctions
in many-variable interacting populations''}
\par\end{center}

\section{Dynamical mean field theory\label{sec:Dynamical-mean-field}}

We derive (3) of the main text using the cavity method, used in many
fields such as mean-field spin glasses \citep{mezard_spin_1987},
neural networks \citep{mezard1989space}, interacting particle systems
in large dimension \citep{liu_dynamics_2021,agoritsas2019out} and
many-variable population dynamics \citep{roy_numerical_2019,bunin_ecological_2017}.
In the many-body dynamics (2) of the main text, each degree of freedom
$x_{i}$ is driven by a `field' $\xi_{i}(t)$

\[
\xi_{i}(t)\equiv\sum_{j}\alpha_{ij}x_{j}(t)\,,
\]
which is expressed as a sum over all the contributions coming from
the other many degrees of freedom it interacts with. Since for any
$i$ the different interaction coefficients $\alpha_{ij}$ are \emph{i.i.d.
}random numbers it is natural to expect that $\xi_{i}(t)$ converges
to a Gaussian process in the large system-size limit $S\to\infty$.
In the cavity method, this is shown by investigating how the dynamics
of a single degree of freedom, say here $x_{0}(t)$, perturbs that
of the ones it is coupled to and by expressing $\xi_{0}(t)$ only
in terms of $x_{i}^{(0)}(t)$ for $i>0$, the evolution of the other
degrees of freedom in an identical system where all couplings to $x_{0}(t)$
would be set to zero. This eventually allows to apply the central
limit theorem and to cast the evolution of $x_{i}(t)$ in the form
of a stochastic differential equation, see (3) of the main text. To
proceed, we start by considering the dynamics of $S$ degrees of freedom
subjected to a perturbation field $h_{i}(t)$ acting as

\[
\dot{x}_{i}=x_{i}(1-x_{i})\left(\sum_{j=1}^{S}\alpha_{ij}x_{j}+h_{i}\right)\,.
\]
For now the field $h_{i}(t)$ is arbitrary but will later represent
the perturbation induced by the coupling to an additional degree of
freedom. The linear response function

\[
R_{ij}(t,s)=\left.\frac{\delta x_{i}(t)}{\delta h_{j}(s)}\right|_{\boldsymbol{h}=0}\,,
\]
obeys

\[
\partial_{t}R_{ij}(t,s)=(1-2x_{i})\left(\sum_{j\neq i}\alpha_{ij}x_{j}\right)R_{ij}(t,s)+x_{i}(1-x_{i})\left(\delta_{ij}\delta(t-s)+\sum_{k\neq i,j}\alpha_{ik}R_{kj}(t,s)+\alpha_{ij}R_{jj}(t,s)\right)\,.
\]
From the above equation it follows that $R_{ii}(t,t)=x_{i}(t)\left[1-x_{i}(t)\right]$
and $R_{ij}(t,t)=0$ for $j\neq i$, from which one can deduce the
scalings $R_{ii}(t,s)=O(1)$ and $R_{ij}(t,s)=\alpha_{ij}\hat{R}_{ij}(t,s)$
with $\hat{R}_{ij}(t,s)=O(1)$ for $i\neq j$. We can now proceed
with the cavity method, by considering a system comprised of $S+1$
degrees of freedom from which we arbitrarily single out one, labeled
$x_{0}$. In the following the indices $i,j$ run from $1$ to $S$.
The dynamics of $x_{0}$ read

\begin{equation}
\dot{x}_{0}=x_{0}(1-x_{0})\sum_{i}\alpha_{0i}x_{i}\,,\label{eq:dyn_x0}
\end{equation}

\noindent and that of the $x_{i}$ read

\begin{equation}
\dot{x}_{i}=x_{i}(1-x_{i})\left(\sum_{j\neq i}\alpha_{ij}x_{j}+\alpha_{i0}x_{0}\right)\,.\label{eq:perturb_x0}
\end{equation}
Therefore a given trajectory of $x_{0}$ acts on the $x_{i}$ as the
previously introduced perturbing field $\boldsymbol{h}$, when setting
$h_{i}=\alpha_{i0}x_{0}$. For a given initial condition $x_{i}(0)$
and a given trajectory $x_{0}(t)$ we decompose the motion of the
$x_{i}(t)$ as $x_{i}(t)=x_{i}^{(0)}(t)+\delta x_{i}[x_{0}](t)$ where
$x_{i}^{(0)}(t)$ is the solution of (\ref{eq:perturb_x0}) when all
the couplings $\alpha_{i0}$ for $i=1\dots S$ are set to zero and
$\delta x_{i}[x_{0}]$ accounts for the correction of the solution
due to the dynamics of $x_{0}.$ It follows from (\ref{eq:dyn_x0})
that describing the dynamics of $x_{0}(t)$ to $O(1)$ in $S$ only
requires to know $\delta x_{i}[x_{0}](t)$ up to order $O(1/\sqrt{S})$.
We can therefore find $\delta x_{i}$ within linear-response, which
up to $O(1/S)$ corrections can be written as

\[
\delta x_{i}[x_{0}](t)=\int_{0}^{t}ds\left(R_{ii}(t,s)\alpha_{i0}+\sum_{j\neq i}R_{ij}(t,s)\alpha_{j0}\right)x_{0}(s)\,.
\]

\noindent Therefore we have to order $O(1)$,

\[
\dot{x}_{0}=x_{0}(1-x_{0})\left[\sum_{i}\alpha_{0i}x_{i}^{(0)}+\int_{0}^{t}ds\sum_{i}\alpha_{0i}\left(R_{ii}(t,s)\alpha_{i0}+\sum_{j\neq i}R_{ij}(t,s)\alpha_{j0}\right)x_{0}(s)\right]\,.
\]

\noindent Because the interaction matrix is fully asymmetric, $\mathbb{E}\left[\alpha_{0j}\alpha_{i0}\right]=0$,
the contribution from the linear response term, $\int_{0}^{t}ds..$,
scales as $O(1/\sqrt{S})$ and can be neglected. The dynamics of $x_{0}$
hence read, up to $O(1/\sqrt{S})$ corrections,
\[
\dot{x}_{0}=x_{0}(1-x_{0})\sum_{i}\alpha_{0i}x_{i}^{(0)}\,.
\]
We now assume that $x_{i}^{(0)}$ and $x_{j}^{(0)}$ (or equivalently
$x_{i}$ and $x_{j}$) are weakly correlated processes for $i\neq j$
meaning that for any functional $F[x(t)]$ we have

\begin{equation}
\mathbb{E}\left[\left(F[x_{i}^{(0)}(t)]-\mathbb{E}[F[x^{(0)}(t)]]\right)\left(F[x_{j}^{(0)}(t)]-\mathbb{E}[F[x^{(0)}(t)]]\right)\right]\underset{S\to\infty}{\to}0\,,\label{eq:weak_correl}
\end{equation}
where we have stressed the fact that all variables are statistically
identical. Such an assumption, which can be verified self-consistently
(see below), is standard in DMFT \citep{krzakala_statistical_2015}.
It implies a law of large numbers, namely that for any functional
$F[x(t)]$

\[
\frac{1}{S}\sum_{i}F[x_{i}^{(0)}(t)]\underset{S\to\infty}{\to}\mathbb{E}[F[x^{(0)}(t)]],
\]
in agreement with the self-averaging property of the auto-correlation
function shown numerically in Fig. 1(D) of the main text. In the large
$S$ limit, $\xi_{i}\equiv\sum_{i}\alpha_{0i}x_{i}^{0}$ thus converges
to a Gaussian process with zero mean and variance

\[
\mathbb{E}\left[\xi_{i}(t)\xi_{i}(t')\right]=\mathbb{E}\left[x^{(0)}(t)x^{(0)}(t')\right]=\mathbb{E}\left[x_{0}(t)x_{0}(t')\right]
\]

\noindent where we used that, up to order $O(1)$, $x_{i}^{(0)}$
and $x_{i}$ and $x_{0}$ are statistically identical. This proves
(3) of the main text. To see that the weak correlation assumption,
(\ref{eq:weak_correl}), is self-consistent within DMFT, we note that
the $O(1)$ dynamics of two degrees of freedom $x_{0}$ and $x_{1}$
read (as the direct interactions between them are only $O\left(S^{-1/2}\right)$)

\[
\dot{x}_{0}=x_{0}(1-x_{0})\sum_{i>1}\alpha_{0i}x_{i}^{(0,1)}\,,
\]

\noindent and

\[
\dot{x}_{1}=x_{1}(1-x_{1})\sum_{i>1}\alpha_{1i}x_{i}^{(0,1)}\,,
\]

\noindent where $x_{i}^{(0,1)}$ refers to the solution of the many-body
dynamics in the absence of both $x_{0}$ and $x_{1}$. Upon assuming
that (\ref{eq:weak_correl}) holds, the moment generating function
of $\xi_{0}=\sum_{i>1}\alpha_{0i}x_{i}^{(0,1)}$ and $\xi_{1}=\sum_{i>1}\alpha_{1i}x_{i}^{(0,1)}$
can be worked out showing that they are independent and identically
distributed Gaussian processes. To leading order, the statistical
independence of $x_{0}$ and $x_{1}$ then follows, in agreement with
(\ref{eq:weak_correl}). This also implies that the exponential growth
rates $\lambda_{i}(t)\equiv\left[1-2x_{i}(t)\right]\xi_{i}(t)$ in
(16) of the main text at the dynamically visited fixed points behave,
to leading order, as independent and identically distributed random
variables in the limit of a large number of degrees of freedom $S\to\infty$.

\section{Lotka-Volterra simulations\label{sec:Lotka-Volterra-simulations}}

In Fig. 1(A) of the main text, the interaction matrix $\mathbf{A}$
is cyclic with $A_{ii}=1,A_{i,i+1}=0.3,A_{i,i-1}=2$, and all $B_{i}=1$.
In Fig. 1(B), $S=2\cdot10^{4}$. The parameters of the Lotka-Volterra
dynamics are $B_{i}=1$ and an interaction matrix $\mathbf{A}$ defined
by $A_{ii}=1$ and $A_{ij}$ for $i\neq j$ Gaussian variables with
mean $\mathbb{E}\left[A_{ij}\right]=10/S$ and variance $\mathbb{E}\left[A_{ij}A_{kl}\right]=2\delta_{ik}\delta_{jl}/S$.

\section{Proof of bounds on $\left\langle f\left(u\left(t\right)\right)f\left(u\left(t'\right)\right)\right\rangle $\label{sec:Bounds-on-the}}

Here we derive the bounds
\[
1/16\le\left\langle f\left(u\left(t\right)\right)f\left(u\left(t'\right)\right)\right\rangle \le1\ ,
\]
stated in the main text, below Eq. (6) there. Indeed, we have first

\[
0<\frac{\text{e}^{u}}{1+\text{e}^{u}}<1\Rightarrow\left\langle f\left(u\left(t\right)\right)f\left(u\left(t'\right)\right)\right\rangle <1\,.
\]
To obtain the lower bound, observe that $u(t)$ and $u(t')$ are jointly
Gaussian with correlation matrix

\[
M(t,t')=\left(\begin{array}{cc}
G(t,t) & G(t,t')\\
G(t,t') & G(t',t')
\end{array}\right)\,.
\]

\noindent Therefore

\begin{align*}
 & \left\langle f\left(u\left(t\right)\right)f\left(u\left(t'\right)\right)\right\rangle >\int_{0}^{+\infty}\int_{0}^{+\infty}\frac{du_{1}}{\sqrt{2\pi}}\frac{du_{2}}{\sqrt{2\pi}}\frac{\exp\left(-\frac{1}{2}\,u^{T}\cdot M^{-1}(t,t')\cdot u\right)}{\sqrt{\det M(t,t')}}f\left(u_{1}\right)f\left(u_{2}\right)\\
 & >\frac{1}{4}\int_{0}^{+\infty}\int_{0}^{+\infty}\frac{du_{1}}{\sqrt{2\pi}}\frac{du_{2}}{\sqrt{2\pi}}\frac{\exp\left(-\frac{1}{2}\,u^{T}\cdot M^{-1}(t,t')\cdot u\right)}{\sqrt{\det M(t,t')}}=\frac{1}{16}\left(1+\frac{2}{\pi}{\rm arccot}\sqrt{\frac{G(t,t)G(t',t')}{G(t,t')^{2}}-1}\,\right)>\frac{1}{16}\ .
\end{align*}
This completes the proof.

\section{Adding a non-zero growth rate\label{sec:Growth-Rate}}

\begin{figure}
\begin{centering}
\includegraphics[width=0.3\columnwidth]{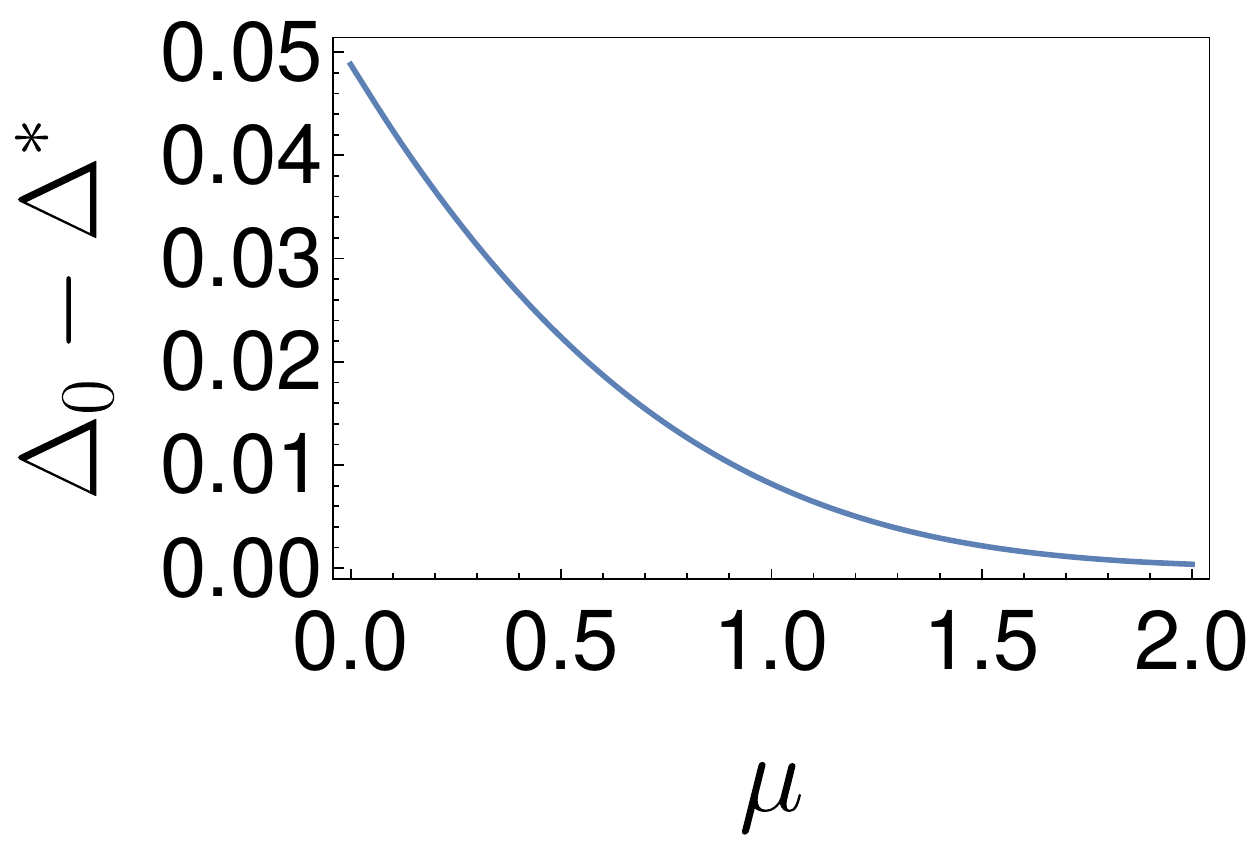}\includegraphics[width=0.3\textwidth]{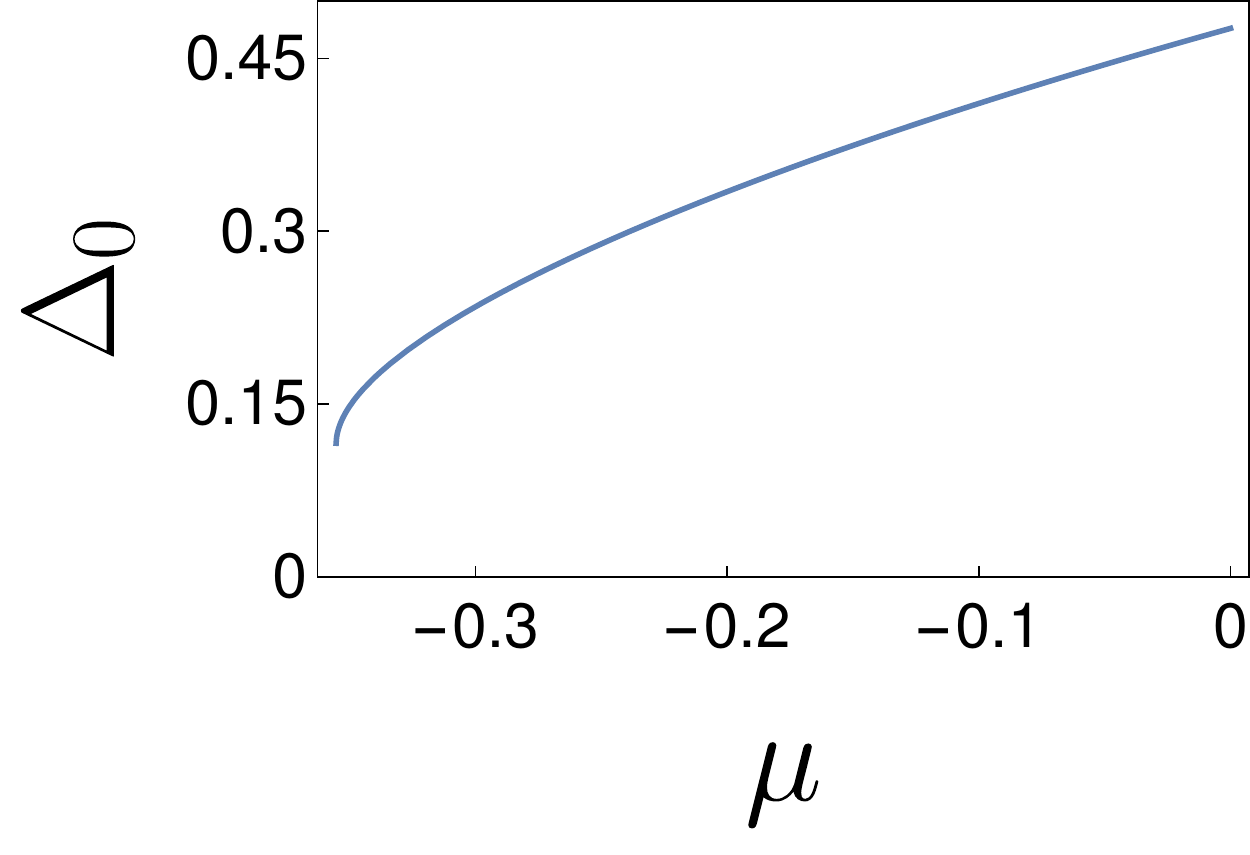}
\par\end{centering}
\caption{\label{fig:g_dep}\textbf{Amplitude of the dynamical fluctuations
of $\tilde{z}(s)$ as a function of the bare growth rate $\mu$ when
all $g_{i}=\mu$.} Left: As $\mu\protect\geq0$ increases, the amplitude
of the temporal fluctuations $\Delta_{0}-\Delta^{*}$ decays continuously
to zero at large $\mu$. Right: For $\mu\le\mu_{c}\approx-0.35$ the
system collapses to the trivial fixed point with all $x_{i}=0$. Shown
is the equal-time correlation function $\Delta_{0}$ as a function
of $\mu$ for $\mu\protect\leq0$.}
\end{figure}

\begin{figure}
\begin{centering}
\includegraphics[width=0.4\textwidth]{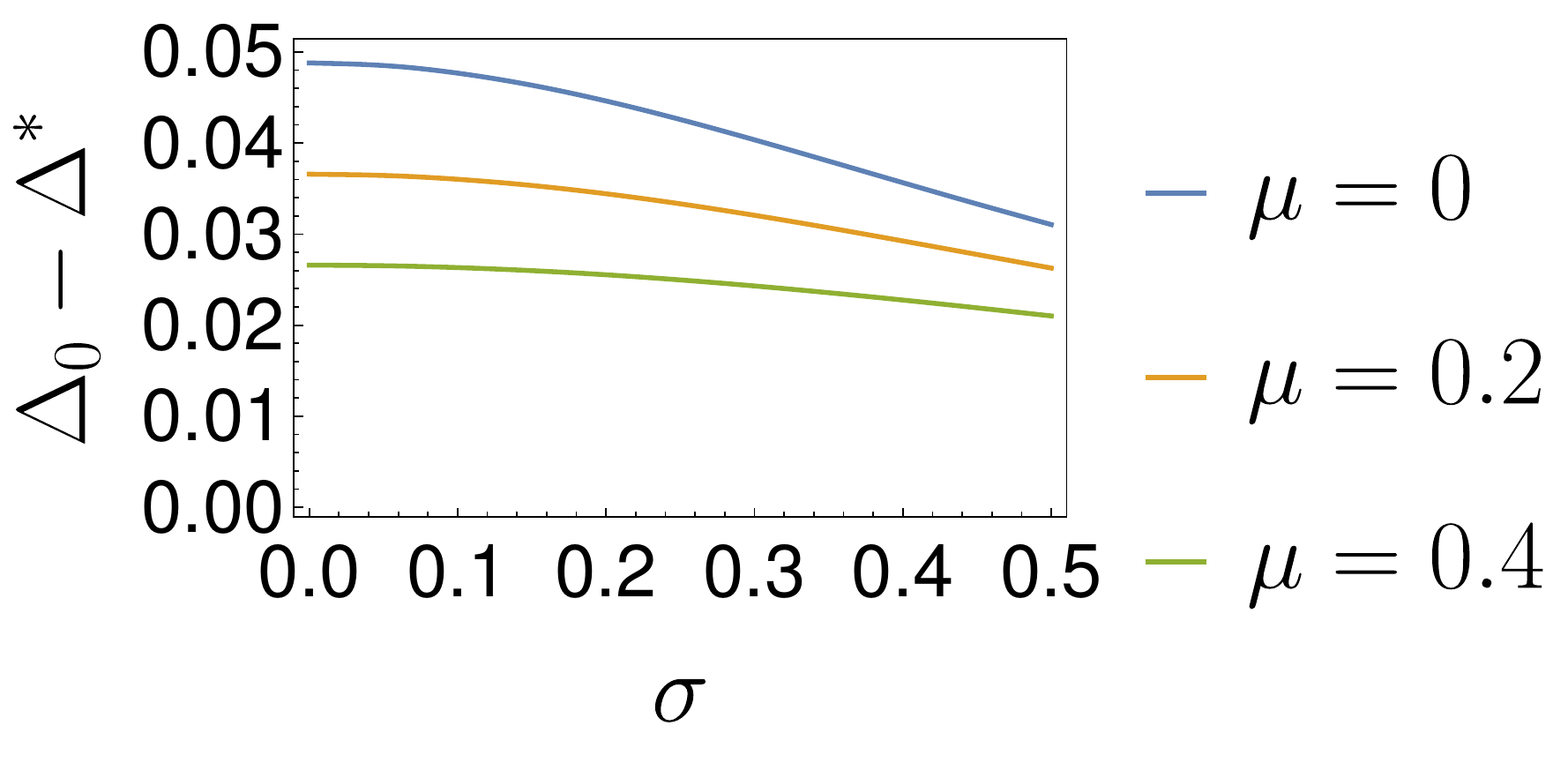}
\par\end{centering}
\caption{\label{fig:g_dep-1}\textbf{Amplitude of the dynamical fluctuations
of $\tilde{z}(s)$ as a function of the standard deviation $\sigma$
of the distribution $P(g)$ for different values of $\mu$. }The amplitude
of the fluctuations $\Delta_{0}-\Delta^{*}$ is a decreasing function
of $\sigma$.}
\end{figure}

The phenomenology presented in the main text can be extended to the
case where a non-zero bare growth rate is taken into account, \emph{i.e.
}for the system of equations
\begin{equation}
\dot{x}_{i}=x_{i}(1-x_{i})\left(g_{i}+\sum_{j=1}^{S}\alpha_{ij}x_{j}\right)\,,\label{eq:many-body-g}
\end{equation}
for $i=1\dots S$ where $g_{i}$ are \emph{i.i.d} species-dependent
growth rates sampled from the distribution $P(g)$ and\textbf{ $\boldsymbol{\alpha}$}
the interaction matrix which is assumed to be Gaussian with zero mean
and variance $\mathbb{E}\left[\alpha_{ij}\alpha_{mn}\right]=\delta_{im}\delta_{jn}/S$.
The analysis follows the one presented in the main text for $g_{i}=0$.
The derivation of the DMFT equations presented in App.~\ref{sec:Dynamical-mean-field}
applies and in the limit $S\to\infty$ the effective stochastic process
reads

\[
\dot{x}=x(1-x)(g+\xi(t))\,,
\]
with $g$ a random variable sampled from $P(g)$ which extends (3)
of the main text and where $\xi(t)$ is a zero-mean Gaussian noise
with variance $\langle\xi(t)\xi(t')\rangle=\langle x(t)x(t')\rangle$
where the average in the right-hand side is now taken over the realisations
of both the noise and the growth rate. If the distribution $P(g)$
has a non-zero support on $[0,+\infty[$, then the bounds of App.~\ref{sec:Bounds-on-the} can be adapted and read
\[
\frac{1}{16}\int_{0}^{+\infty}dg\,P(g)\le\left\langle \xi(t)\xi(t')\right\rangle \le1\,.
\]
Accordingly, by defining $s=\ln t$ and $z=\ln\left[x/(1-x)\right]/t$,
(7) of the main text becomes
\begin{equation}
z'(s)=-z(s)+g+\hat{\xi}(s)\,,\label{eq:DMFT_g}
\end{equation}
with (8) holding in the long-time limit. We introduce $\tilde{z}(s)=z(s)-g$
and denote its autocorrelation function $\left\langle \tilde{z}(s)\tilde{z}(s')\right\rangle \equiv\tilde{\Delta}(s-s')$
which is related to that of the process $z(s)$ by $\left\langle \tilde{z}(s)\tilde{z}(s')\right\rangle \equiv\left\langle z(s)z(s')\right\rangle -\left\langle g^{2}\right\rangle $.
The equation for the evolution of $\tilde{\Delta}''(s)$ then follows
from 
\[
\tilde{\Delta}''(s)=-V'(\tilde{\Delta},\Delta_{0})\,,
\]
where $\Delta_{0}=\tilde{\Delta}(0)$ and the effective potential
$V(\tilde{\Delta},\Delta_{0})$ reads
\[
V\equiv-\frac{\tilde{\Delta}^{2}}{2}+\frac{\tilde{\Delta}}{4}\left(1+\int_{-\infty}^{+\infty}dg\,P(g)\,\text{Erf}\left(\frac{g}{\sqrt{2\Delta_{0}}}\right)\right)+\int_{-\infty}^{+\infty}dg\,P(g)\int_{0}^{+\infty}\frac{dx}{2\sqrt{2\pi\Delta_{0}}}\text{e}^{-\frac{(x-g)^{2}}{2\Delta_{0}}}\int_{0}^{\tilde{\Delta}}d\Delta\,\text{Erf}\left(\frac{g(\Delta_{0}-\Delta)+x\Delta}{\sqrt{2\Delta_{0}(\Delta_{0}^{2}-\Delta^{2})}}\right).
\]
Following the discussion of the main text, we find $\Delta_{0}$ and
$\Delta^{*}=\lim_{s\to\infty}\tilde{\Delta}(s)$ by requiring that
$V(\Delta^{*},\Delta_{0})=V(\Delta_{0},\Delta_{0})$ together with
$V'(\Delta^{*},\Delta_{0})=0$. We first consider the case $P(g)=\delta(g-\mu)$,
corresponding to an identical bare growth rate for all the species.
If $\mu\geq0$, the behavior is similar to the $g=0$ case studied
at depth in the main text. The noise splits into two independent contributions:
one static and one with time-translation invariant statistics in $\log$-time.
The amplitude of the temporal fluctuations of the process $\tilde{z}(s)$
varies continuously with the bare growth rate $\mu$ and decays to
zero at large $\mu$, see Fig. \ref{fig:g_dep}. All the degrees of
freedom in (2) are indeed expected to reach the limit $x_{i}(t)\underset{t\to\infty}{\to}1$
and dynamical fluctuations to be suppressed in this limit. If $\mu\leq0$,
numerical solutions of the equations for $\Delta_{0}$ and $\Delta^{*}$
suggest that there exists a finite value $\mu_{c}\leq0$ such that
for strong enough negative rate $\mu\leq\mu_{c}$, we have $\Delta_{0}=\Delta^{*}=0$
corresponding to the trivial fixed point where all species are extinct,
$x_{i}=0$, see Fig. \ref{fig:g_dep}. We also investigated the case
where there is heterogeneity in the growth rates by taking $P(g)$
Gaussian with mean $\mu$ and standard deviation $\sigma$. The results
remain qualitatively the same, with the amplitude of the fluctuations
decreasing with $\sigma$, see Fig. \ref{fig:g_dep-1}.

The derivation can be extanded to account for the existence of species-dependent
growth rates $g_{i}$, in the case where the $g_{i=1\dots S}$ are
identically distributed and sampled independently from each other
and from the elements of the interaction matrix \textbf{$\boldsymbol{\alpha}$
}from a distribution $P(g)$\textbf{.}

\section{Ergodic measure for $x(s)$\label{sec:Ergodic-measure-for}}

We recall (7) of the main text

\begin{equation}
z'(s)=-z(s)+\hat{\xi}(s)\,,\label{eq:dynz}
\end{equation}

\noindent and decompose the noise $\hat{\xi}(s)$ as 
\[
\hat{\xi}(s)=\bar{\xi}+\delta\xi(s)
\]

\noindent with $\bar{\xi}$ a Gaussian random variable with zero mean
and variance $\left\langle \bar{\xi}^{2}\right\rangle =\Delta^{*}$
and $\delta\xi(s)$ an independent Gaussian process with zero mean
and covariance \emph{
\[
\left\langle \delta\xi(s)\delta\xi(s')\right\rangle =\hat{C}(s,s')-\Delta^{*}\,.
\]
}

\noindent In the long-time limit, the solution to (\ref{eq:dynz})
reads

\[
z(s)=\bar{\xi}+\text{e}^{-s}\int_{0}^{s}ds'\text{\,e}^{-s'}\delta\xi(s')\,,
\]
so that at large $s$, and fixed $\bar{\xi}$, $z(s)$ is a Gaussian
variable with mean $\bar{\xi}$ and variance 
\[
\left\langle (z(s)-\bar{\xi})^{2}\right\rangle \underset{s\to\infty}{\to}\int du\text{e}^{-u}\left[\hat{C}(u)-\Delta^{*}\right]\,.
\]
Equation (15) of the main text then follows,
\begin{equation}
P_{\bar{\xi}}(x)=(1-p)\,\delta(x)+p\,\delta(x-1)\,,\label{eq:ergodic_measure-1}
\end{equation}
with $p=\left[1+\text{Erf}\left(\bar{\xi}/\sqrt{2\chi}\right)\right]/2$,
where $\bar{\xi}$ is a zero mean Gaussian random variable with variance
$\left\langle \bar{\xi}^{2}\right\rangle =\Delta^{*}$ and $\chi=\int_{0}^{\infty}ds\text{e}^{-s}\left[\hat{C}(s)-\Delta^{*}\right]$.
While each variable switches between 0 and 1 an infinite amount of
time, the probability distribution of $p$ (the fraction, in log-time,
spent at $x=1$) diverges at 0 (and accordingly at 1) as,
\[
P(p)\underset{p\to0}{\sim}\left(p\sqrt{-\ln p}\right)^{-1+\frac{\chi}{\sqrt{\Delta^{*}}}}\,,
\]
with $\chi/\sqrt{\Delta^{*}}\simeq0.92$. Namely, some degrees of
freedom are strongly biased towards one of the boundaries.

\section{Stability spectrum\label{sec:Stability-spectrum}}

As stated in the main text, the joint distribution of $\xi(t)$ and
$u(t)$ is Gaussian,

\begin{equation}
P_{t}(u,\xi)=\frac{1}{2\pi\sqrt{\det\boldsymbol{H}}}\exp\left[-\frac{1}{2}(u,\xi)\cdot\boldsymbol{H}^{-1}\cdot(u,\xi)\right]\,,\label{eq:double_gaussian-1}
\end{equation}
 with the matrix $\boldsymbol{H}$ given by

\[
\boldsymbol{H}=\left(\begin{array}{cc}
G(t,t) & \left\langle u(t)\xi(t)\right\rangle \\
\left\langle u(t)\xi(t)\right\rangle  & C(t,t)
\end{array}\right)\,.
\]
Using this equation, changing variables to $\left(u,\lambda\right)$
and integrating over $u$, the probability distribution of $\lambda(t)$
is found to be
\begin{align}
\rho_{t}(\lambda)= & \frac{\sqrt{1+\kappa^{2}}}{\kappa\sqrt{\pi C(t,t)}}\int_{0}^{+\infty}\frac{du}{\sqrt{\pi}}\exp\left[-\frac{1+\kappa(t)^{2}}{2\kappa(t)^{2}}u^{2}-\frac{1+\kappa(t)^{2}}{2\kappa(t)^{2}}\frac{\lambda^{2}}{C(t,t)}\left(\frac{1-\text{e}^{\sqrt{G(t,t)}u}}{1+\text{e}^{\sqrt{G(t,t)}u}}\right)^{2}+\frac{\sqrt{1+\kappa(t)^{2}}}{\kappa(t)^{2}\sqrt{C(t,t)}}u\lambda\left(\frac{1-\text{e}^{\sqrt{G(t,t)}u}}{1+\text{e}^{\sqrt{G(t,t)}u}}\right)^{2}\right]\nonumber \\
 & \times\left|\frac{1-\text{e}^{\sqrt{G(t,t)}u}}{1+\text{e}^{\sqrt{G(t,t)}u}}\right|\,,\label{eq:full_measure_lambda}
\end{align}
with

\[
\kappa(t)=\sqrt{\frac{C(t,t)G(t,t)}{\left\langle u(t)\xi(t)\right\rangle ^{2}}-1}\,.
\]
At long times,

\begin{align*}
\kappa(t)\to\kappa_{\infty} & =\lim_{t\to\infty}\sqrt{\frac{\Delta_{0}}{2\left(\int_{0}^{1}C(ts,t)\,ds\right)^{2}}-1}=\sqrt{\frac{1}{2\Delta_{0}}-1}\simeq0.224\,.
\end{align*}

\noindent where the last equality was obtained by noting that
\begin{align*}
\lim_{t\to\infty}\int_{0}^{1}C(ts,t)\,ds & =\int_{0}^{+\infty}ds\,\text{e}^{-s}\hat{C}(s)=\int_{0}^{+\infty}ds\,\text{e}^{-s}\left[-\hat{\Delta}''(s)+\hat{\Delta}(s)\right]=-\int_{0}^{+\infty}ds\,\frac{d}{ds}\left\{ \text{e}^{-s}\left[\hat{\Delta}'(s)+\hat{\Delta}(s)\right]\right\} =\Delta_{0}\,.
\end{align*}
At $t\rightarrow\infty$, equation (\ref{eq:full_measure_lambda})
reduces to (17) of the main text.

\end{document}